\documentclass{JFM-FLM_Au}

\newcommand{\textin}[1]{\mbox{\scriptsize{#1}}}

\definecolor{grisclair}{rgb}{0.6,0.6,0.6}

\lefttitle{M. Rubio, S. Rodríguez-Aparicio, J. M. Montanero and M. A. Herrada}
\righttitle{Journal of Fluid Mechanics}

\title{Interface pinch-off in the presence of a soluble surfactant}

\author{\aff{1}M. Rubio, \aff{2}S. Rodríguez-Aparicio, \aff{2}J. M. Montanero \and \aff{3}M. A. Herrada}

\affiliation{\aff{1}Depto.\ de Ingenier\'{\i}a Energ\'etica y Fluidomec\'anica and Instituto de las Tecnolog\'{\i}as Avanzadas de la Producci\'on (ITAP), Universidad de Valladolid, E-47003 Valladolid, Spain
\aff{2}Depto.\ de Ingenier\'{\i}a Mec\'anica, Energ\'etica y de los Materiales and Instituto de Computaci\'on Cient\'{\i}fica Avanzada (ICCAEx), Universidad de Extremadura, E-06006 Badajoz, Spain
\aff{3}Departamento de Ingenier\'{\i}a Aeroespacial y Mec\'anica de Fluidos, Universidad de Sevilla, E-41092 Sevilla, Spain}

\corresau{J. M. Montanero, \email{jmm@unex.es}}

\begin{document}
\maketitle

\begin{abstract}
We study numerically and experimentally the breakup of a pendant droplet loaded with a soluble surfactant. We consider the limit in which surfactant sorption is limited only by diffusion. Surfactant transfer toward the interface is enhanced by convection. As a consequence, diffusion does not constitute a significant barrier over most of the breakup, and surfactant sorption maintains the surface tension practically constant across the interface.  Diffusion hinders the surfactant sorption only very close to the interface pinch-off. The droplet shape in the diffusion-limited model deviates significantly from that in the insoluble case over most of the breakup. In the insoluble case, the droplet shape is affected by surfactant depletion, which leads to a local increase in surface tension and Marangoni stress. The dynamics of a millimeter-sized droplet loaded with Surfynol 465 agree remarkably well with predictions from the diffusion-limited model, without any parameter fitting, down to pinching times of the order of $10-20$ $\mu$s. Sodium dodecyl sulfate (SDS) produces essentially the same effects as those for Surfynol 465. Therefore, both Surfynol 465 and SDS maintain a practically constant surface tension throughout most of the droplet breakup. Slow-kinetics surfactants, such as Triton X-100, differ significantly from Surfynol 465 and SDS. The most evident effect of the surfactant adsorption energy barrier is the shortening of the filament that bridges the upper meniscus and the detached lower drop. Comparing the filament length to that of a clean interface with the same surface tension allows one to evaluate the rate of surfactant adsorption.
\end{abstract}

\maketitle

\section{Introduction}
\label{int}

% Surfactants
The use of surfactants has numerous applications across diverse technological fields, such as ink-jet printing, the food industry, and biotechnology. For instance, surfactants enhance the stability of emulsions and bubbles, reduce surface tension to adjust wettability, regulate transport across interfaces, and enable the encapsulation of cells and biomolecules in drops \citep{A16}. The presence of a surfactant monolayer significantly affects the dynamical response of capillary systems even at large scales.

% Modeling surfactants
Modeling surfactants in fluid dynamics is challenging \citep{MS20,M24}. The surfactant molecules dissolved in the liquid are convected and diffused across the hydrodynamic bulk. They accumulate in the sublayer adjacent to the interface, which serves as a source/sink of surfactant molecules during system evolution. The net adsorption/desorption flux can be obtained from a kinetic model, which typically involves unknown adsorption and desorption constants. The surfactant molecules are also convected and diffused over the interface. The surfactant surface density is affected by sorption kinetics, interface surfactant transport, and the dilatation/compression of the interface element. The uneven surfactant distribution at the interface leads to local soluto-capillarity (a local reduction in surface tension) and to Marangoni stress, which significant affect the fluid dynamics at small length scales. 

% droplet breakup
In the pinch-off of a Newtonian clean interface, the system spontaneously approaches a finite-time singularity, offering a unique opportunity to observe fluid behavior at arbitrarily small length and time scales. This property makes this problem an ideal candidate for questioning our knowledge of fundamental aspects of fluid dynamics \citep{PSS90,E93,P95,LS16,KWTB18,CMP09,VMHF14,CCTSHHLB15,PMHVV17,DHHVRKEB18}. The breakup of a pendant droplet is probably the simplest problem in which an interface pinches.

% Results for insoluble surfactants
The effect of a surfactant on droplet breakup has been studied assuming the surfactant is insoluble \citep{RABK09,PMHVV17,KWTB18,PRHEM20,MS20b}. This hypothesis is based on the fact that the characteristic times for surfactant sorption and/or diffusion are much longer than the breakup time of sufficiently small droplets. The noticeable effect of local solutocapillarity \citep{RABK09,SPADBK12}, Marangoni stress \citep{SL90b,AB99,TL02,CMP02,MB06,HSYLBP08,JGS06,LFB06,DSXCS06,HM16b,KWTB18}, and surfactant viscosity \citep{PMHVV17,PRHEM20} has been described under that assumption. 

% Solubility
Considering surfactant solubility considerably increases the complexity of the problem for two reasons: (i) the sorption kinetics is difficult to model, especially for non-dilute systems, and (ii) the large Péclet number characterizing diffusion across the bulk hinders the calculation of the surfactant boundary layer at the interface. This probably explains why this fundamental problem has not yet been tackled.

% Expected solubility effects
During the first phase of the breakup of a pendant drop, a thin liquid filament forms between the liquid attached to the feeding capillary and the dripping drop. The size and shape of that filament can be affected by surfactant solubility due to the lower speed of this part of the breakup process, especially for a fast-kinetics surfactant, for which sorption kinetics are presumably much faster than this phase of the droplet breakup \citep{JGS06,KJMS18}. One may expect the droplet shape to lie between that of an insoluble surfactant and that of a droplet with a constant and uniform coverage of surfactant, as occurs in other configurations \citep{ML94b}.  

Both surfactant convection and the expansion/compression rate of the interface element diverge in the interface neck as the interface approaches breakup. For this reason, it is natural to hypothesize that surfactant diffusion towards the interface becomes subdominant sufficiently close to the pinch-off point. However, the radial size of the pinching region vanishes as the interface approaches breakup, leading to a divergent surfactant diffusion flux. Therefore, it is not evident whether and when diffusion becomes irrelevant as the neck interface radius vanishes.   

% This paper
In this paper, we numerically examine the effect of surfactant solubility on the breakup of a millimeter-sized water drop. We consider a fast-kinetics surfactant, assuming that the surfactant sublayer and the interface are at local equilibrium at all times, which implies that sorption kinetics is much faster than any other process in droplet dynamics. This assumption allows us to describe sorption kinetics solely in terms of the Langmuir adsorption isotherm, which involves the depletion length and maximum packing fraction, measured from equilibrium surface tension values. Under the fast-kinetics surfactant approximation, the exchange of surfactant between the bulk and the interface is limited by diffusion, allowing us to elucidate its effect on droplet breakup. We conduct experiments with surfactants exhibiting different sorption kinetics rates (Surfynol 465, sodium dodecyl sulfate (SDS), and Triton X-100) to assess the extent to which sorption kinetics constitutes a significant energy barrier during the droplet breakup. We determine the major effect of the adsorption energy barrier on pendant droplet breakup, thereby establishing a simple test bench to assess the rate at which a surfactant adsorbs onto the interface.

% Structure of the paper
The paper is organized as follows. Section \ref{for} formulates the problem in dimensionless terms. The governing equations, the limits considered in our analysis, and the numerical method are described in Sec.\ \ref{sec3}. Section \ref{exp} shows the experimental method. The numerical and experimental results are presented in Secs.\ \ref{num} and \ref{exp2}, respectively. Finally, the paper closes with some concluding remarks in Sec.\ \ref{conclusions}. 

\section{Formulation of the problem}
\label{for}

% Clean
Consider a droplet of a liquid with density $\rho$, viscosity $\mu$, and surface tension $\hat{\sigma}_0$ hanging from a capillary of radius $R$ under the action of the gravity $g$. The droplet is formed by injecting liquid through the capillary under quasi-static conditions and breaks up when its volume exceeds its maximum value. The dimensionless droplet shape ${\cal S}$ at the maximum volume stability limit is a function only of the Bond number $B=\rho g R^2/\hat{\sigma}_0$.

\begin{figure}
\begin{center}
\resizebox{0.25\textwidth}{!}{\includegraphics{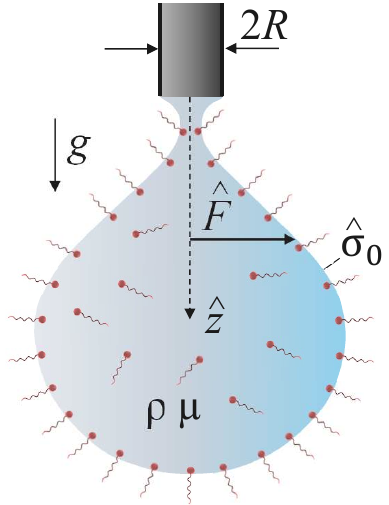}}
\end{center}
\caption{Sketch of the fluid configuration.}
\label{numer0}
\end{figure}

We will pay special attention to the evolution of the interface contour $\hat{F}(\hat{z},\hat{t})$, where $\hat{F}$ is the distance of the interface element from the symmetry axis, $\hat{z}$ is the axial coordinate ($\hat{z}=0$ corresponds to the position of the triple contact line), and $\hat{t}$ is the dimensional time. The dimensionless interface contour $F=\hat{F}/R$ is a function of the initial droplet shape ${\cal S}$, the Bond number $B$, the Ohnesorge number Oh$=\mu/(\rho R \hat{\sigma}_0)^{1/2}$, and the dimensionless axial coordinate $z=\hat{z}/R$ and time $t=\hat{t}/t_c$, where $t_{c}=(\rho R^3/\hat{\sigma}_0)^{1/2}$ is the inertio-capillary time. The Bond number affects the interface evolution through both the initial droplet shape and the gravitational force acting during the droplet breakup, i.e., 
\begin{equation}
\label{e1}
F=F({\cal S}(B);B,\text{Oh};z,t).
\end{equation}
For sufficiently small Bond numbers, $F=F({\cal S}(B);\text{Oh};z,t)$. The effect of $B$ through ${\cal S}$ cannot be neglected, no matter how small $B$ is.

% Surfactant. Insoluble
In the presence of an {\em insoluble} surfactant, one also considers the initial surface coverage $\Gamma_0=\hat{\Gamma}_0/\hat{\Gamma}_{\infty}$, where $\hat{\Gamma}_0$ is the initial surfactant surface concentration and $\hat{\Gamma}_{\infty}$ is the maximum packing concentration, as well as the Marangoni number Ma=$\hat{\Gamma}_{\infty} R_g T/\hat{\sigma}_0$ and the surface Péclet number $\text{Pe}_s=R v_c/{\cal D}_{s0}$ number. Here, $R_g$ and $T$ are the gas constant and temperature, respectively, $\hat{\sigma}_0$ is the initial (equilibrium) surface tension, $v_c=R/t_{c}$ is the inertio-capillary velocity, and ${\cal D}_{s0}$ is the surface diffusion coefficient in the dilute limit $\Gamma_0\to 0$. Therefore,
\begin{equation}
F=F({\cal S}(B);B,\text{Oh};\Gamma_0,\text{Ma},\text{Pe}_s;z,t).
\end{equation}

% Surfactant. Soluble
In the presence of a {\em soluble} surfactant, the sorption kinetics are typically characterized by the Biot number $\text{Bi}=k_a/v_c$ ($k_a$ is the adsorption constant) and dimensionless depletion length $\Lambda_d=L_d/R$ ($L_d=k_a/k_d$ is the depletion length and $k_d$ the desorption constant) (see Sec.\ \ref{sec3}). The transport of surfactant across the bulk is described in terms of the Péclet number $\text{Pe}=R v_c/{\cal D}$, where ${\cal D}$ is the volumetric diffusion coefficient. In this case,
\begin{equation}
\label{com}
F=F({\cal S}(B);B,\text{Oh};\Gamma_0,\text{Ma},\text{Pe}_s;\text{Bi},\Lambda_d,\text{Pe};z,t).
\end{equation}

% Fast-kinetics surfactant
Equation (\ref{com}) reflects the complexity of the breakup of a pendant droplet containing a soluble surfactant in the general case. If sorption kinetics are much faster than any other process, the surfactant sublayer and the interface are at equilibrium. In this case, the problem becomes independent of $\text{Bi}$ (see Sec.\ \ref{limits}), i.e.,
\begin{equation}
F=F({\cal S}(B);B,\text{Oh};\Gamma_0,\text{Ma},\text{Pe}_s;\Lambda_d,\text{Pe};z,t).
\end{equation}
Apart from the diffusion coefficients, the surfactant parameters entering the problem are the depletion length $L_d$ and the maximum packing concentration $\hat{\Gamma}_{\infty}$, which can be determined from the surfactant equilibrium isotherm. We will conduct numerical simulations in the fast-kinetics (diffusion-limited) limit. The results will be compared with experimental data for Surfynol 465 to assess how well this approximation captures droplet breakup. The solubility effects will be analyzed by comparison with the insoluble limit.

\section{Governing equations and numerical method}
\label{sec3}

\subsection{Governing equations}

% Bulk equations
This section describes the general governing equations. The limits considered in this work are explained in Sec.\ \ref{limits}. We consider $R$, $t_{c}=(\rho R^3/\hat{\sigma}_0)^{1/2}$, $v_c=R/t_c$, $\hat{\sigma}_0/R$, and $\hat{\Gamma}_{\infty}/R$ as the characteristic length, time, velocity, pressure, and surfactant concentration, respectively. The dimensionless Navier-Stokes equations for the axisymmetric velocity $\mathbf{v}(r,z;t)=u(r,z;t) \vec{u}_r+v(r,z;t)\vec{u}_z$ and reduced pressure $p(r,z;t)$ fields are
\begin{equation}
\left(ru\right)_r+rw_z=0, \label{basic1}
\end{equation}
\begin{eqnarray}
\frac{\partial u}{\partial t} + u u_r+ w u_z=-p_r+\text{Oh}\left(u_{rr}+(u/r)_r+u_{zz}\right),\nonumber\\ \label{basic2}
\end{eqnarray}
\begin{eqnarray}
\frac{\partial w}{\partial t} + u w_r+ ww_z=-p_z+\text{Oh}\left(w_{rr}+w_r/r+w_{zz}\right),\nonumber \\
\label{basic3}
\end{eqnarray}
where the subscripts $r$ and $z$ denote the partial derivatives with respect to the corresponding coordinates. 

% Interface boundary conditions
The kinematic compatibility at the interface yields
\begin{equation}
\frac{\partial F}{\partial t}+F_z w-u=0.\label{int1}
\end{equation}
at $r=F(z,t)$. We neglect the effect of the outer gas. The equilibrium of normal stresses on that surface leads to
\begin{equation}
\label{n}
-p+B\, z+\tau_n=-\sigma{\boldsymbol \nabla}_s\cdot {\bf n},
\end{equation}
where \begin{equation}
\tau_n=\text{Oh}\frac{2[u_r-F_z(w_r+u_z)+F_z^{2}w_z]}{1+F_z^{2}},
\end{equation}
is the normal viscous stress, $\sigma=\hat{\sigma}/\hat{\sigma}_0$ is the local value $\hat{\sigma}$ of the surface tension normalized by its initial value $\hat{\sigma}_0$, ${\boldsymbol\nabla_s}$ is the tangential intrinsic gradient along the interface ($s$ is the intrinsic coordinate along the interface), and ${\bf n}$ is the unit outward normal vector. 
The equilibrium of tangential stresses yields
\begin{eqnarray}
\label{t}
\tau_t=\tau^{\textin{Ma}}, \label{int2}
\end{eqnarray}
where $\tau_t$ is the tangential viscous stress
\begin{equation}
\label{t1}
\tau_t=\text{Oh}\left[(1-F_z^{2})(w_r+u_z)+2F_z(u_r-w_z)\right],
\end{equation}
and $\tau^{\textin{Ma}}$ is the Marangoni stress
\begin{equation}
\label{t2}
\tau^{\textin{Ma}}=\sigma_z(1+F_z^{2})^{1/2}.
\end{equation}
The viscous surface stresses have been neglected in Eqs.\ (\ref{n}) and (\ref{t}) because, for the surfactant considered in our analysis, they become relevant only very close to the interface pinching \citep{PRHEM20}.

% Surfactants 
We will restrict our analysis to surfactant concentrations below the critical micelle concentration. The monomer volumetric concentration (measured in terms of the $\hat{\Gamma}_{\infty}/R$) $c(r,z;t)$ is calculated from the conservation equation \citep{CMP09,KB19}
\begin{equation}
\label{c11}
\frac{\partial c}{\partial t}+u c_r+w c_z=\text{Pe}^{-1}\left[(r c_{r})_r/r+c_{zz}\right].
\end{equation}

The net sorption flux ${\cal J}$ at the interface is calculated as \citep{CMP09,HYSL15,KB19}
\begin{equation}
\label{flux}
{\cal J}={\cal J}_a-{\cal J}_d,\quad\quad {\cal J}_a=\text{Bi}\, c_s(1-\Gamma),\quad \quad {\cal J}_d=\text{Bi}\, \Lambda_d^{-1}\, \Gamma,
\end{equation}
where ${\cal J}_a=\hat{{\cal J}}_aR/(v_c\hat{\Gamma}_{\infty})$ and ${\cal J}_d=\hat{{\cal J}}_dR/(v_c\hat{\Gamma}_{\infty})$ are the dimensionless adsorption and desorption fluxes, respectively, $\hat{{\cal J}}_a$ and $\hat{{\cal J}}_d$ are their dimensional counterparts, and $c_s$ is the surfactant concentration evaluated at the interface. The net sorption flux equals the surfactant diffused from/to the bulk ${\cal J}_{\cal D}$; i.e.,
\begin{equation}
\label{mio}
{\cal J}={\cal J}_{\cal D}=\text{Pe}^{-1}\left. \nabla c\right|_n=\left.\frac{c_z F_z-c_r}{\text{Pe}\sqrt{1+F_z^2}}\right|_{r=F}.    
\end{equation}
This equation couples surfactant transport across the bulk and over the interface.

The surfactant surface concentration $\Gamma$ verifies the advection-diffusion equation \citep{CMP09}
\begin{equation}
\label{ad}
\frac{\partial \Gamma}{\partial t}+{\boldsymbol \nabla}_s\cdot(\Gamma {\bf v}_s)+\Gamma {\bf n}\cdot ({\boldsymbol \nabla}_s\cdot {\bf n}){\bf v}=\frac{1}{\text{Pe}_s}{\boldsymbol \nabla}_s\cdot \left(\frac{{\boldsymbol \nabla}_s\Gamma}{1-\Gamma}\right)+{\cal J},
\end{equation}
where ${\bf v}_s={\boldsymbol {\sf I}}_s{\bf v}=v_s {\bf t}$ is the (two-dimensional) surface velocity, ${\bf t}$ is the tangential unit vector, ${\boldsymbol {\sf I}}_s={\boldsymbol {\sf I}}-{\bf n}{\bf n}$ is the tensor that projects any vector on that surface, being ${\boldsymbol {\sf I}}$ is the identity tensor. It is worth mentioning that the diffusion coefficient has been calculated as  
\begin{equation}
{\cal D_S}=\frac{{\cal D}_{S0}}{1-\Gamma},
\end{equation}
where ${\cal D}_{S0}$ is its value in the dilute Henry limit $\Gamma\to 0$ \citep{MS20}. Considering a constant diffusion coefficient outside that limit is thermodynamically inconsistent \citep{MS20}.

The dependence of the surface tension $\sigma$ upon the surfactant surface concentration $\Gamma$ is calculated from the Langmuir equation of state \citep{T97}
\begin{equation}
\sigma=1+\text{Ma}\ \log\left(\frac{1-\Gamma}{1-\Gamma_0}\right).
\label{es}
\end{equation}

% Boundary conditions
The anchorage condition of the triple contact line, $F=1$, is imposed at the edge of the feeding capillary. The numerical integration of (\ref{ad}) is performed considering zero surfactant diffusive flux at the triple contact line. We consider the standard regularity conditions $u=w_r=c_r=0$ at the symmetry axis.

% Initial conditions
The following initial condition is used in our simulation. At $t=0$, we consider a droplet in equilibrium under a slightly reduced gravitational force $(1-\beta)B$, with $0<\beta\ll 1$. We prescribe the droplet volume 
\begin{equation}
{\cal V}=\pi\int_{0}^{z_a} F^2\ dz\label{volume}
\end{equation}
corresponding to equilibrium for that reduced gravitational force ($z_a$ is the axial coordinate of the droplet apex). Notice that this volume is slightly larger than the maximum volume at the stability limit for the Bond number value $B$. We also prescribe the surfactant surface concentration $\Gamma_0$ and the volume concentration $c_0(\Gamma_0)$ given by the Langmuir adsorption isotherm
\begin{equation}
\label{km20}
\frac{\Gamma_0}{1-\Gamma_0}=\Lambda_d c_0.
\end{equation}

For $t>0$, we set the Bond number value to $B$, so that the droplet becomes unstable. The droplet evolves under that condition for the rest of the simulation. For sufficiently small values of the Bond number step $\beta$, the droplet dynamics corresponds to that in the experiment (see Sec.\ \ref{exp}).

\subsection{Limiting cases}
\label{limits}

In the insolubility limit, the exchange of surfactant molecules between the bulk and the interface is much slower than the droplet breakup. Then, ${\cal J}=0$ in Eq.\ (\ref{ad}), and the surfactant transport across the bulk does not need to be solved.

If sorption kinetics are much faster than any other process, the surfactant sublayer and the interface are at local equilibrium at all times, and ${\cal J}_a={\cal J}_d$. This last condition leads to the Langmuir adsorption isotherm
\begin{equation}
\label{km}
\frac{\Gamma}{1-\Gamma}=\Lambda_d c_s,
\end{equation}
which is prescribed as a Dirichlet boundary condition when calculating $c(r,z;t)$.   This ``fast kinetics" limit corresponds to $\text{Bi}\to\infty$ and $[\Lambda_d c_s-\Gamma/(1-\Gamma)]\to 0$ so that 
\begin{equation}
\text{Pe}^{-1}\left. \nabla c\right|_n=\frac{\text{Bi}(1-\Gamma)}{\Lambda_d}\left(\Lambda_d c_s-\frac{\Gamma}{1-\Gamma}\right)
\end{equation}
remains finite \citep{WSB14}. In this limit, ${\cal J}$ is replaced by ${\cal J}_{\cal D}$ in Eq.\ (\ref{ad}), implying that surfactant transfer to/from the interface is controlled only by diffusion. 

Finally, if both surfactant kinetics and diffusion are much faster than any other process, $c(r,z;t)=c_0$, $\Gamma(s;t)=\Gamma_0$, and therefore $\sigma=1$ over the interface. This can be regarded as the opposite limit to the insoluble one. For this reason, we will refer to it as the {\em perfectly soluble} case. This limit corresponds to a clean interface with the same surface tension as that of the surfactant-loaded droplet at equilibrium. 

We will compare the numerical results obtained in the insoluble case (${\cal J}=0$), in the diffusion-limited (fast-kinetics) limit (${\cal J}_a={\cal J}_d$), and the perfectly soluble case ($\Gamma=\Gamma_0$). 

\subsection{Numerical method}
\label{nume}

% Spatial and temporal discretization
The theoretical model is numerically solved using a variant of the method proposed by \citet{HM16a} (see also \citet{HERRADA25}). The time-dependent liquid region is mapped onto a fixed numerical domain through a coordinate transformation. The transformed spatial domains were discretized using $n_{\eta}=41$ Chebyshev spectral collocation points in the transformed radial direction and $n_{\xi}=4001$ equally spaced collocation points in the transformed axial direction. The axial direction was discretized using fourth-order finite differences. Second-order backward finite differences were used to discretize the time domain. The time-integration method is implicit and adaptive. The step size is chosen by comparing the norm of the difference between solutions obtained using first- and second-order finite differences. If the norm exceeds a threshold $\epsilon_1$, the time step $\Delta t$ is halved. If it is greater than another $\epsilon_2$ ($\epsilon_2<\epsilon_1$), $\Delta t$ is multiplied by 1.5. To handle the interface overturning occurring just before droplet breakup, a quasi-elliptic transformation \cite{DT03} was applied to generate the mesh. 

% Breakup time
The time-dependent mapping of the physical domain prevents the algorithm from surpassing the free-surface pinch-off. The breakup time, $t_b$, in the simulation was calculated from the linear extrapolation of the last $N_b=20$ values of the minimum interface radius $F_{\textin{min}}(t)$. We checked that the value of $N_b$ does not significantly affect the results.

\begin{figure}
\begin{center}
\resizebox{0.45\linewidth}{!}{\includegraphics{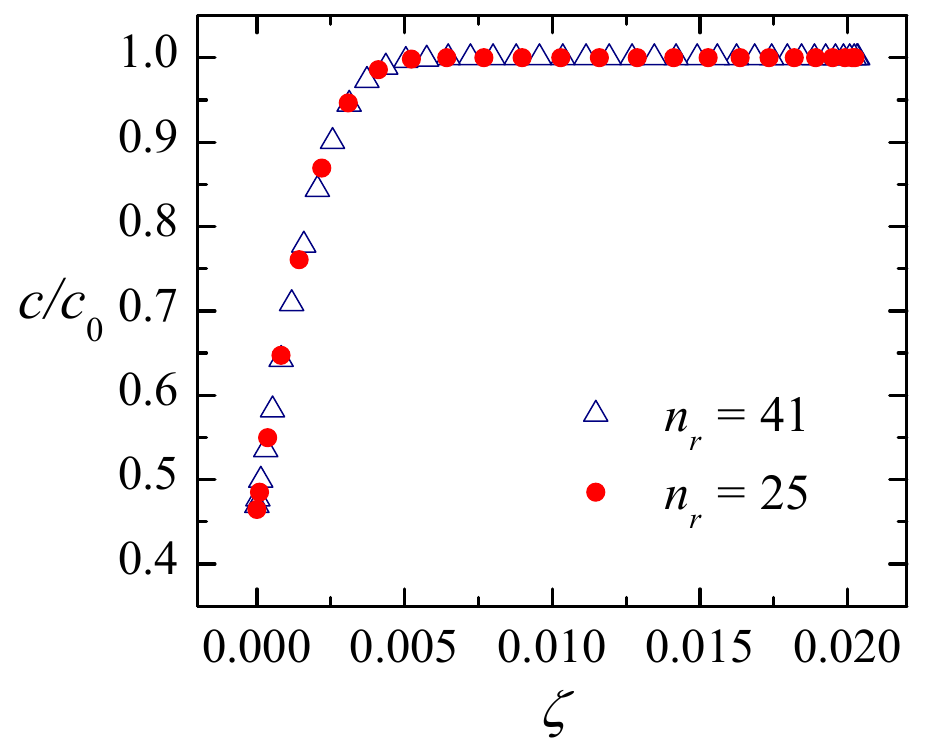}}
\end{center}
\caption{Surfactant volumetric concentration $c/c_0$ as a function of the distance $\zeta$ to the interface in the perpendicular direction. The concentration was evaluated at the interface neck and corresponds to the instant when $F_{\textin{min}}=0.02$. The results were calculated in the diffusion-limited (fast-kinetics) limit for $\beta=0.005$, $B=0.0954$, $\text{Oh}=6.18\times 10^{-3}$, $\Gamma_0=0.997$, $\Lambda_d=0.599$, $\text{Ma}=0.125$, $\text{Pe}_s=1.62\times 10^{4}$, and $\text{Pe}=1.62\times 10^{4}$.}
\label{boun}
\end{figure}

% Test diffusion
Our numerical method enables accurate resolution of the surfactant boundary layer near the interface, even at the high Péclet number characteristic of this problem. Figure \ref{boun} shows the surfactant volumetric concentration $c$ at the interface neck ($F=F_{\textin{min}}$) as a function of the distance $\xi$ to the interface. The results do not depend significantly on the number of grid points in the direction perpendicular to the interface, despite the boundary layer thickness, $\delta\simeq 0.0025$, being three orders of magnitude smaller than the capillary radius. 

% Diffusion time
Consider an initially clean planar surface that remains at rest in a liquid bath containing a low-concentration surfactant. When adsorption is controlled by diffusion, the surfactant surface concentration obeys the equation \citep{JBS04,AWA10,MS20}
\begin{equation}
\frac{\Gamma}{\Gamma_{\textin{eq}}}=1-e^{t/t_D}\, \text{erfc}\left(\sqrt{t/t_D}\right),
\end{equation}
where $\Gamma_{\textin{eq}}$ is the equilibrium value, and $t_D=\text{Pe}\, \Lambda_d^2$ is the (dimensionless) characteristic diffusion time. For the case analyzed in Fig.\ \ref{boun}, $t_D\sim 10^4$. The droplet breakup occurs at dimensionless times of order unity (the breakup time is of order the inertio-capillary time). This indicates that diffusion alone cannot transport surfactant toward the interface. However, the boundary layer thickness $\delta$ can be two orders of magnitude smaller than the depletion length $\Lambda_d=0.599$ (Fig.\ \ref{boun}). If we consider $\delta$ instead of $\Lambda_d$ to estimate the diffusion time ($t_D=\text{Pe}\, \delta^2$), then $t_D\sim 1$. This indicates that the thinning of the liquid filament considerably enhances surfactant diffusion near the pinching point, thereby affecting the dynamics of the surfactant monolayer. We will return to this point in Sec.\ \ref{num}. 

\section{Experimental method}
\label{exp}

\begin{figure}[ht]
\begin{center}
\includegraphics[width=7.5cm]{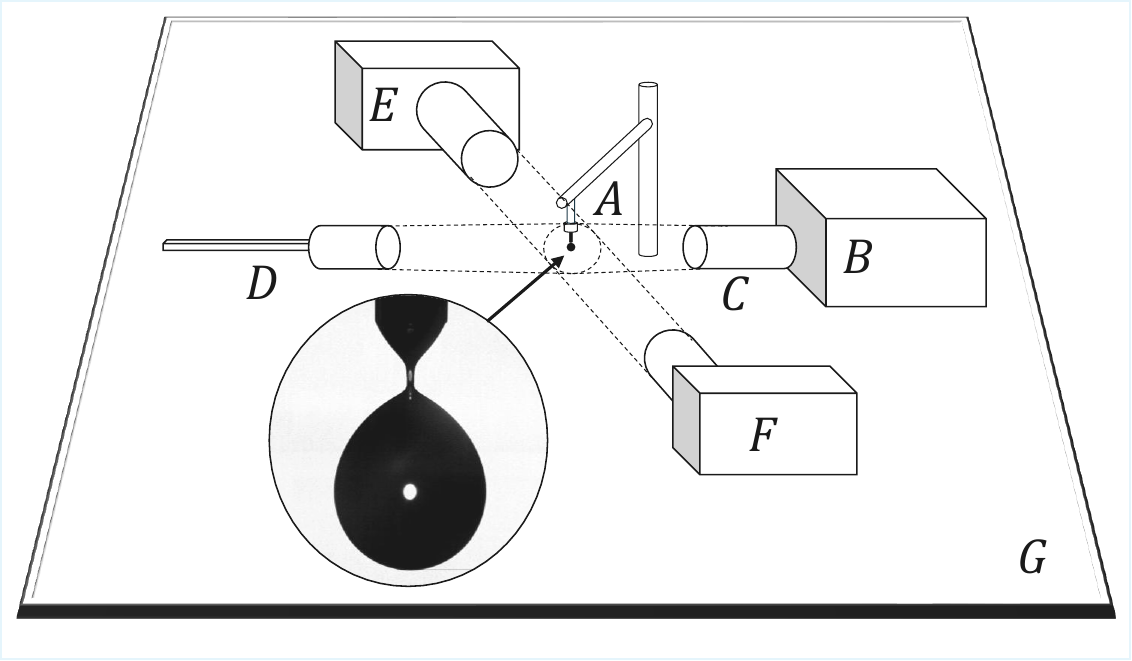}
\end{center}
\caption{Experimental setup: feeding capillary (A), ultra-high speed video camera (B), optical lenses (C), laser (D), optical trigger (E),  cold-white backlight (F) and anti-vibration isolation system (G).}
\label{sketch}
\end{figure}

% Experimental setup
The experimental method is similar to that used by \citet{PRHEM20} to study the breakup of a submillimeter droplet. In the experimental setup (Fig.\ \ref{sketch}), a cylindrical feeding capillary (A) with an outer radius $R=0.635$ mm  was placed vertically. A pendant droplet was formed quasi-statically by injecting the liquid at 0.2 ml/h using a syringe pump (Harvard Apparatus PHD 4400) driven by a stepping motor. The triple contact lines were pinned to the capillary's outer edge. When this limit was reached, the droplet spontaneously broke up.

% Images
Digital images of the droplet were taken using an ultra-high-speed video camera ({\sc Kirana}-5M) (B) equipped with optical lenses (an Optem HR 50X magnification zoom-objective and a NAVITAR 12X set of lenses) (C)  (Fig.\ \ref{sketch}). The images were acquired at speeds in the range $2 \times 10^{4}-2 \times 10^{6}$ fps and with magnifications in the range $1.95-7.79$ $\mu$m/pixel. Both the speed and magnification were adjusted to visualize different stages of droplet breakup. The camera was illuminated by a laser (SI-LUX 640, {\sc Specialised Imaging}) (D) synchronized with the camera, reducing the effective exposure time to 100 ns. The camera was triggered by an optical trigger (SI-OT3, {\sc Specialised Imaging}) (E), equipped with optical lenses and illuminated by a cold-white backlight (F). All these elements were mounted on an optical table with a pneumatic anti-vibration isolation system (G) to damp the vibrations coming from the building. The images of droplet breakup were analyzed using a subpixel-resolution method \citep{M24} to determine the free-surface location.

\begin{figure}
\begin{center}
\resizebox{0.5\linewidth}{!}{\includegraphics{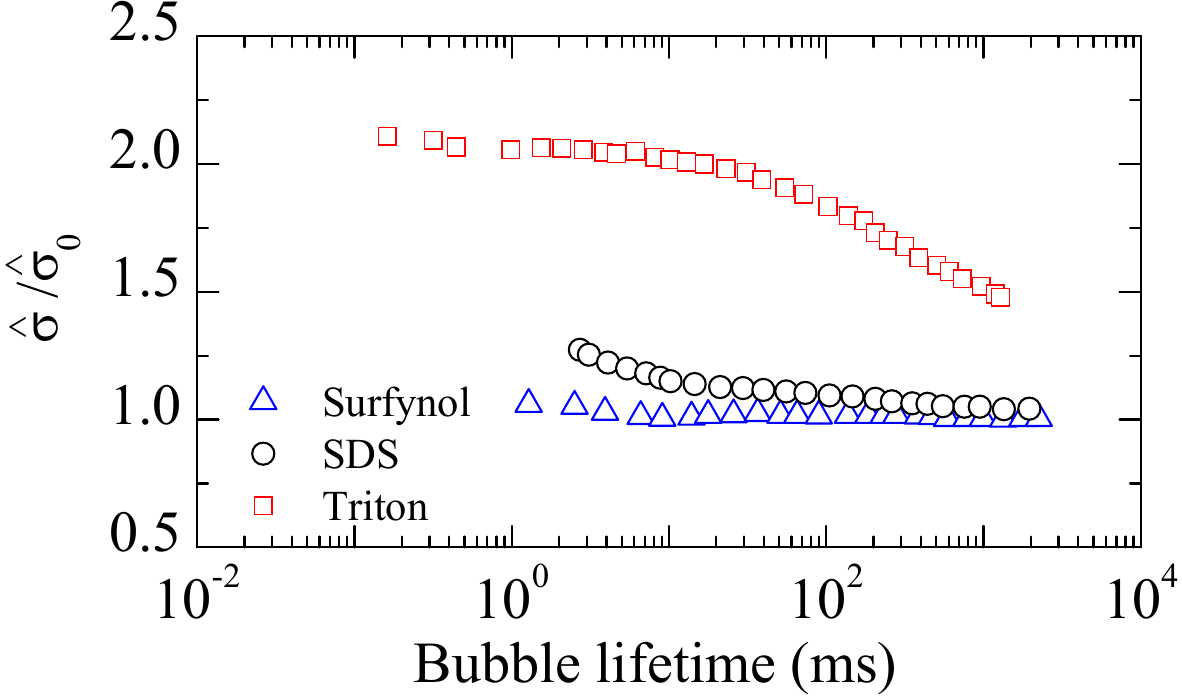}}
\end{center}
\caption{Dynamic surface tension of aqueous solutions of Surfynol 465 at $\hat{c}/\hat{c}_{\text{cmc}}=1.3$, SDS at $\hat{c}/\hat{c}_{\text{cmc}}=1.3$, and Triton X-100 at $\hat{c}/\hat{c}_{\text{cmc}}=1$. The values were normalized with the corresponding equilibrium values $\hat{\sigma}_0=29.2$ mN/m, 35.8 mN/m, and 34.2 mN/m for Surfynol 465, SDS, and Triton X-100, respectively. The data was measured by \citet{VSQCC24} using the maximum bubble-pressure tensiometer.}
\label{bmp}
\end{figure}

% Surfynol 465
We conducted experiments with Surfynol 465 (Evonik) dissolved in ultra-pure Milli-Q water at 2.1 mol/m$^3$. The physical properties of this solution were $\rho=997$ kg/m$^3$, $\mu=1.0$ mPa$\cdot$s, and $\hat{\sigma}_0=41.3$ mN/m. The Surfynol 465 sorption constants have not yet been measured. However, \citet{VSQCC24} showed that the adsorption rate of Surfynol 465 is higher than that of other surfactants at the same relative concentration $\hat{c}/\hat{c}_{\text{cmc}}$ (Fig.\ \ref{bmp}), where $\hat{c}_{\text{cmc}}$ is the critical micelle concentration. In this sense, Surfynol 465 is a fast-kinetics surfactant. The bulk and surface diffusion coefficients of Surfynol 465 have not been determined experimentally either. One expects these coefficients to take values in the range $10^{-8}-10^{-10}$ m$^2$/s, as occurs to most surfactants with similar molecular weights \citep{T97}. The results presented in Sec.\ \ref{num} were calculated for ${\cal D}={\cal D}_{s0}=10^{-8}$ m$^2$/s. 

% SDS and Triton
Additionally, we conducted experiments with SDS and Triton X-100 (Sigma-Aldrich) at 6.034 and 0.059 mol/m$^3$, respectively. We selected SDS in water because it is widely used and well characterized. The droplet breakup occurs on a time scale given by a few inertio-capillary times $t_{c}\simeq 2.5$ ms, i.e., of the order of 10 ms. The results obtained by the bubble pressure method suggest that a significant transfer of SDS from the bulk to the interface occurs on that timescale \citep{VSQCC24}. Therefore, SDS cannot be considered as an insoluble surfactant in our experiments. This approximation yields accurate results only for much smaller values of the feeding capillary radius $R$ \citep{PRHEM20}. Conversely, SDS is expected to behave as a fast-kinetic surfactant during most of the millimeter droplet breakup. The adsorption characteristic time of Triton X-100 is supposed to be significantly longer than that of SDS (Fig.\ \ref{bmp}). We expect significant deviations of the experimental results from the diffusion-limited model prediction in this case.

\begin{figure*}
\begin{center}
\resizebox{0.5\linewidth}{!}{\includegraphics{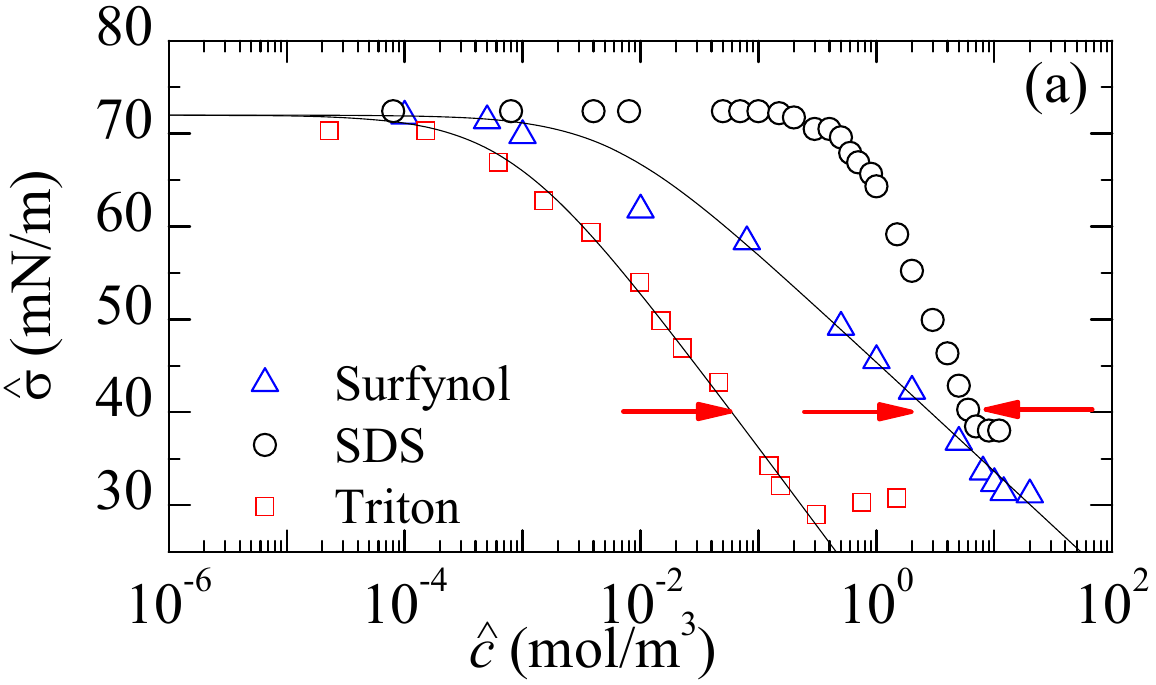}}\resizebox{0.47\linewidth}{!}{\includegraphics{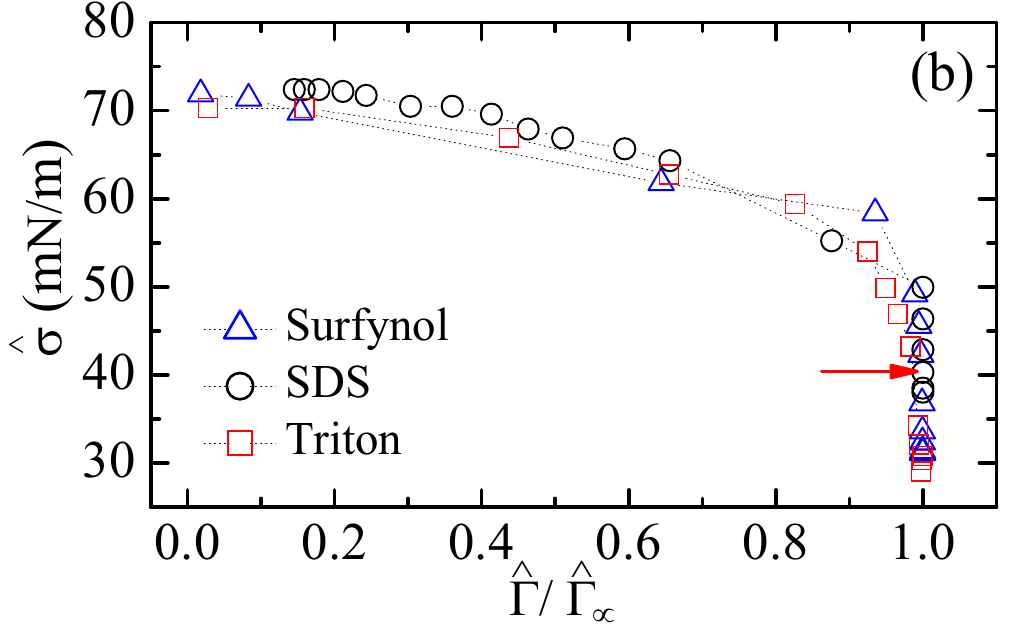}}
\end{center}
\caption{(a) Surface tension $\hat{\sigma}$ as a function of the surfactant volumetric concentration $\hat{c}$. The lines are the fits of the Langmuir equation of state to the Surfynol 465 and Triton X-100 data. (b) Surface tension $\hat{\sigma}$ as a function of the normalized surfactant surface concentration $\hat{\Gamma}/\hat{\Gamma}_{\infty}$. The arrows indicate the concentrations considered in the experiments.}
\label{st}
\end{figure*}

% Surface tensions. Surfynol 465 and Triton
Figure \ref{st}a shows the surface tension $\hat{\sigma}$ as a function of the surfactant volumetric concentration $\hat{c}$ for Surfynol 465, SDS, and Triton X-100. The Langmuir equation of state
\begin{equation}
\hat{\sigma}=\hat{\sigma}_c-\hat{\Gamma}_{\infty} R_g T\log(1+L_d\, \hat{c}/\hat{\Gamma}_{\infty})
\end{equation}
was fitted to the Surfynol 465 and Triton X-100 data. Here, $\hat{\sigma}_c=72$ mN/m is the water-air surface tension. In the case of Surfynol 465, the value $\hat{\Gamma}_{\infty}=2.1$ $\mu$mol/m$^2$ was taken from \citet{PZPA16}, and $L_d=0.38$ mm was obtained from the fit. In the case of Triton X-100, the values $\hat{\Gamma}_{\infty}=3.03$ $\mu$mol/m$^2$ and $L_d=3.83$ mm were obtained from the fit. Using these values, the Langmuir adsorption isotherm (\ref{km})
\begin{equation}
\label{km2}
\frac{\hat{\Gamma}/\hat{\Gamma}_{\infty}}{1-\hat{\Gamma}/\hat{\Gamma}_{\infty}}=\frac{L_d\,  \hat{c}_s}{\hat{\Gamma}_{\infty}}
\end{equation}
allows one to obtain $\hat{\sigma}(\hat{\Gamma})$ from $\hat{\sigma}(\hat{c}_s)$ (Fig.\ \ref{st}b). The arrows in  Fig.\ \ref{st} indicate the concentrations considered in the experiments. The critical micelle concentrations for Surfynol 465, SDS, and Triton X-100 are $\hat{c}_{\text{cmc}}=10$, 8.1, and 0.24 mM, respectively, higher than the concentrations considered in our experiment.

% Surface tension. SDS
The values of $\hat{\sigma}(\hat{c})$ and $\hat{\sigma}(\hat{\Gamma}/\hat{\Gamma}_{\infty})$ for SDS were taken from \citet{TMS70}. The values $\hat{\Gamma}_{\infty}=3.9$ $\mu$mol/m$^2$ and $L_d=0.037$ mm were taken from \citet{TMS70} too. It is worth noting that rigorous work should employ equations of state $\hat{\sigma}(\hat{\Gamma})$ derived specifically for ionic, electrostatically interacting monolayers to describe the equilibrium surface tension in the presence of SDS. In practice, the Langmuir equation of state usually folds electrostatic effects into effective parameters \citep{PF01}. This is empirically convenient and can reasonably predict the surface density at the pure water-air interface \citep{PF01}, but should not be interpreted as SDS truly verifying the ideal Langmuir equation of state over the full isotherm. For this reason, we will not use the model described in Sec.\ \ref{sec3} to simulate the breakup of a droplet loaded with SDS.

\section{Numerical results}
\label{num}

% Choices
This section presents numerical results for the three limits described in Sec.\ \ref{limits}: insoluble, diffusion-limited, and perfectly-soluble. We will consider the parameters values $B=0.0954$, $\text{Oh}=6.18\times 10^{-3}$, $\Gamma_0=0.997$, $\Lambda_d=1.67$, $\text{Ma}=0.125$, $\text{Pe}_s=1.62\times 10^{4}$, and $\text{Pe}=1.62\times 10^{4}$, which correspond to droplet of an aqueous solution of Surfynol 465 at the surfactant concentration $\hat{c}/\hat{c}_{\textin{cmc}}=0.21$ hanging on a capillary $R=0.635$ mm in outer radius (see Sec.\ \ref{exp}). The interface expands during the droplet breakup, reducing the surfactant surface concentration. We verified that the surface tension given by the Langmuir equation of state (\ref{es}) took realistic values across the entire interface at all times. 

% The phases
We can distinguish three phases of the droplet breakup. First, the droplet essentially remains at equilibrium while a very small-amplitude perturbation grows on a timescale $\omega_i^{-1}$, where $\omega_i$ is the growth rate of the linear mode triggering the instability. Second, a thin liquid filament connecting the upper and lower liquid volumes is formed. This non-linear phase occurs on a timescale of the order of the inertio-capillary time. Then, the filament neck thins on a much shorter timescale until the interface pinches. We will study these three phases of droplet breakup separately.

\subsection{Droplet instability}

\begin{figure*}
\begin{center}
\resizebox{0.48\linewidth}{!}{\includegraphics{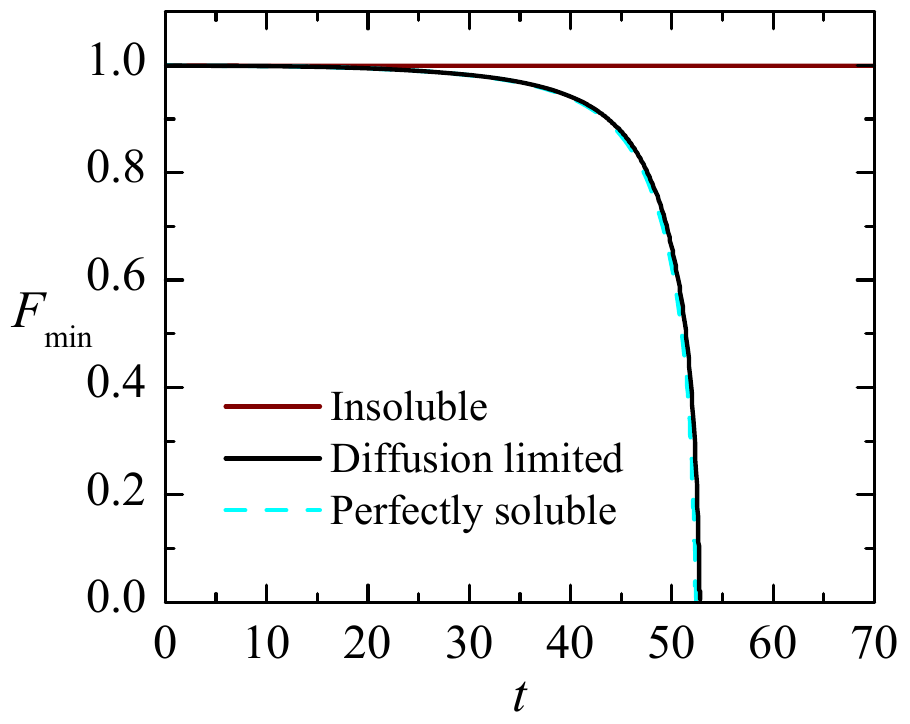}}\resizebox{0.5\linewidth}{!}{\includegraphics{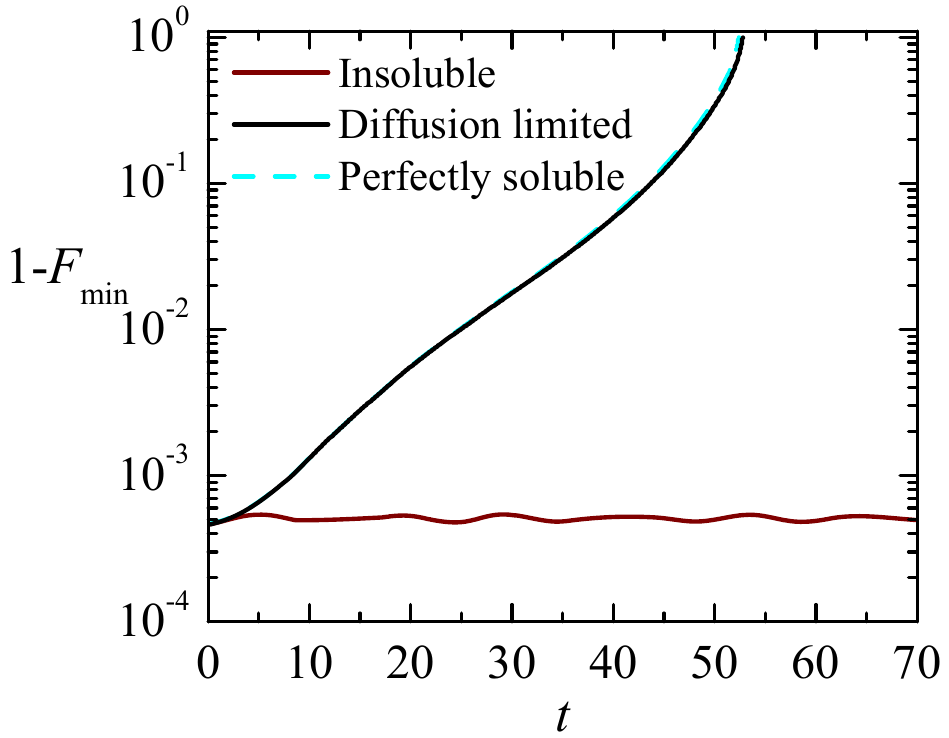}}
\end{center}
\caption{$F_{\textin{min}}$ as a function of the time $t$. The origin of time $t=0$ corresponds to the instant at which the step of the Bond number was introduced.}
\label{st0}
\end{figure*}

% Diffusion-limited
We begin our analysis by considering a tiny Bond number step $\beta=0.005$ to trigger the droplet instability (see Sec.\ \ref{nume}). Under the diffusion-limited approximation, surfactant adsorption effectively compensates for surfactant depletion during the initial phase of droplet breakup, preventing the formation of surfactant gradients at the interface and, consequently, eliminating the stabilizing effect of the Marangoni stress. In fact, the growth rate of the critical capillary mode that leads to pendant-droplet breakup is practically the same as that of the perfectly soluble case (a clean interface with the same equilibrium surface tension). This can be observed in Fig.\ \ref{st0}, which shows the minimum (neck) interface radius $F_{\textin{min}}$ as a function of time. 

% Insoluble
Interestingly, the droplet does not break up during the simulated time for the insoluble case (Fig.\ \ref{st0}). During the perturbation growth, the interface stretches, the surfactant redistributes, and surface tension gradients develop. These gradients generate Marangoni stress, which can suppress tangential motion (interface immobilization) and reduce the instability mode's growth rate. As a consequence, the droplet does not evolve toward the breakup over the time analyzed in the simulation. A similar effect has been described under different conditions \citep{KWCCAB20,WKLB24,CBKSCJM21}.

% Comment
It is worth noting that the surfactant affects only the development of the instability, not the stability limit. The critical mode at the pendant droplet detachment is the same as the neutral mode of the equilibrium Young–Laplace operator. The stabilizing Marangoni stress does not enter the Young-Laplace equation, and therefore, the surfactant does not shift the maximum volume stability limit. 

% Link
To compare the droplet breakups described by the insoluble, diffusion-limited, and perfectly soluble models, we increased the Bond number step size $\beta$ from $\beta=0.005$ to 0.2 so that the droplet broke up during the computed time in the insoluble case as well. We now focus on the last two nonlinear phases of the droplet breakup: (i) the formation of a thin liquid filament connecting the upper and lower liquid volumes on times of the order of the inertio-capillary time, and (ii) the much faster thinning of the interface neck until the interface breakup. Section \ref{fil} and \ref{bre} show results for the first and second phases mentioned above, respectively. 

\subsection{Formation of the liquid filament}
\label{fil}

\begin{figure}
\begin{center}
\resizebox{0.55\linewidth}{!}{\includegraphics{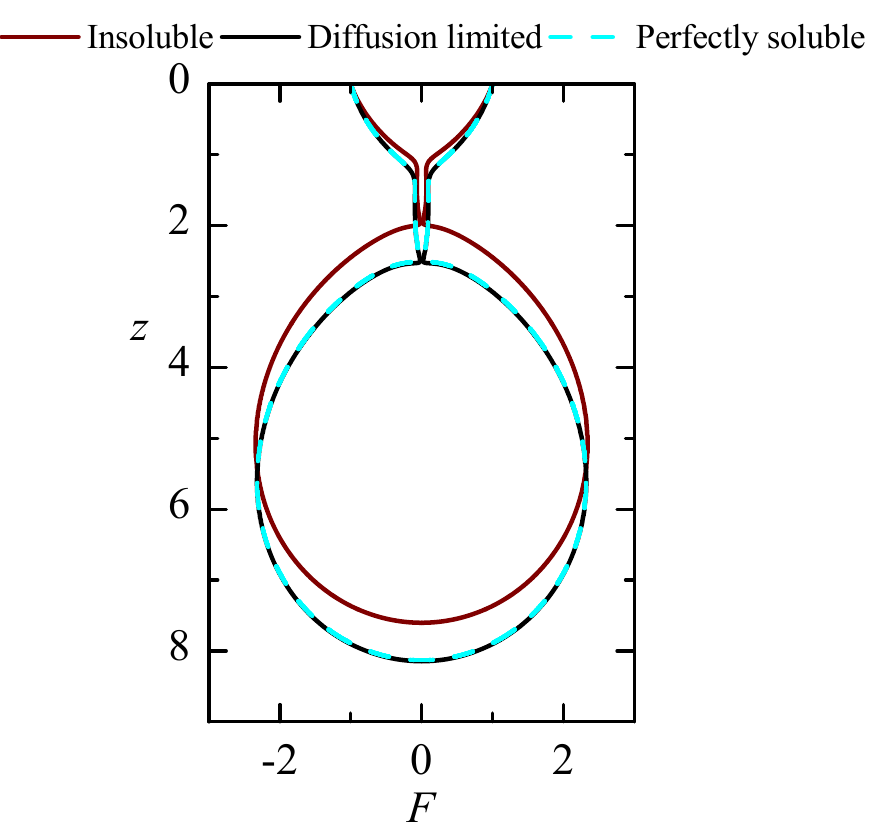}}
\end{center}
\caption{Interface contour at the instant for which $F_{\textin{min}}=0.02$.}
\label{st6}
\end{figure}

\begin{figure}
\begin{center}
\resizebox{0.6\linewidth}{!}{\includegraphics{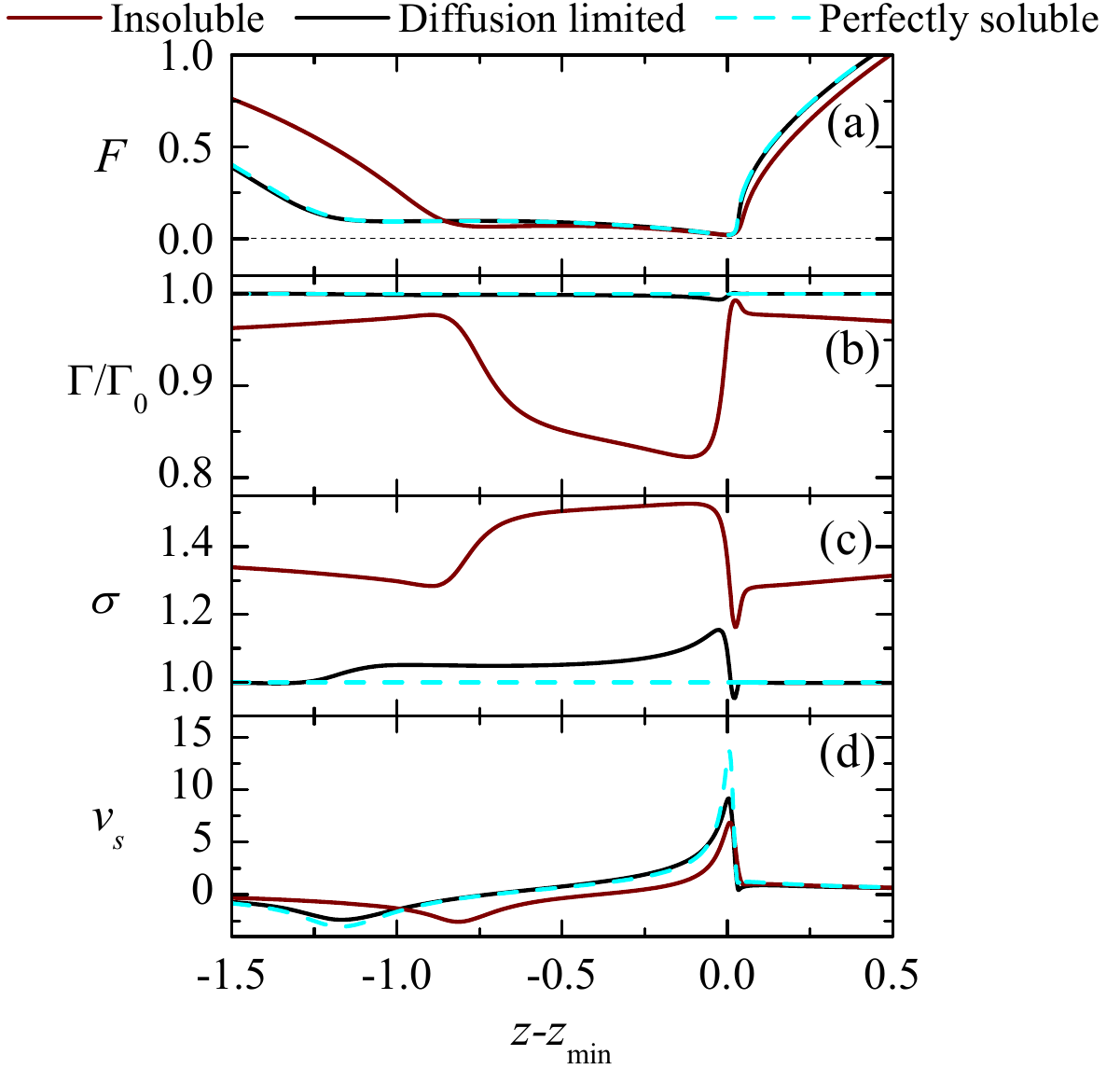}}
\end{center}
\caption{Interface contour $F$ (a), surfactant surface concentration $\Gamma/\Gamma_0$ (b), surface tension $\sigma$ (c), and tangential surface velocity $v_s$ (d) at the instant for which $F_{\textin{min}}=0.02$ ($z_{\textin{min}}$ is the vertical coordinate of the interface neck).}
\label{st7}
\end{figure}

% Diffusion-limited
Figure \ref{st6} compares the shapes of the liquid filament connecting the upper and lower liquid volumes calculated with the three models. We consider the shapes for the same value of $F_{\textin{min}}$ to eliminate the influence of the different growth rates of the unstable linear mode leading to breakup in the three cases. The diffusion-limited and perfectly soluble models predict practically the same filament shape, indicating that diffusion manages to keep $\Gamma\simeq \Gamma_0$ at least for $F_{\textin{min}}\geq 0.02$ (Fig.\ \ref{st7}b). As explained in Sec.\ \ref{nume}, this occurs because the filament thinning sharpens the surfactant boundary layer (Fig.\ \ref{boun}), enhancing surfactant diffusion towards the interface. There is a tiny reduction of $\Gamma$ in the diffusion-limited case, which produces an increase in $\sigma$ of the order of 10\% (Fig.\ \ref{st7}c). This occurs due to the sharp dependence of $\sigma$ on $\Gamma$ for $\Gamma\simeq 1$ (Fig.\ \ref{st}b). However, this effect does not significantly alter the filament shape. The resulting Marangoni stress slightly immobilizes the interface with respect to the perfectly soluble (clean interface) case near $z=z_{\textin{min}}$ (Fig.\ \ref{st7}d). 

% Insoluble
The insoluble approximation yields a significantly shorter filament (Fig.\ \ref{st6}a). In the upper part of the filament, the interface radius $F(z)$ in the insoluble case is larger than in the other two cases due to the significant increase in the surface tension (Fig.\ \ref{st7}c). This increase is caused by surfactant depletion and the fact that $\sigma$ sharply depends on $\Gamma$ form $\Gamma\simeq 1$ (Fig.\ \ref{st}b). This is the most noticeable footprint of the surfactant adsorption energy barrier in the droplet breakup. In addition, Marangoni stress reduces the tangential surface velocity $v_s$ over practically the entire filament (Fig.\ \ref{st7}d).  

\begin{figure}
\begin{center}
\resizebox{0.5\linewidth}{!}{\includegraphics{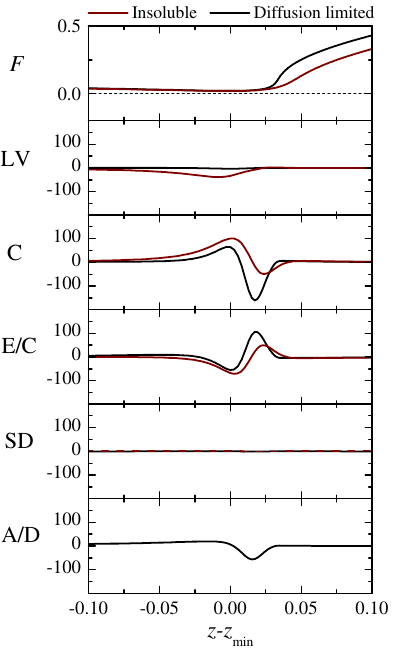}}
\end{center}
\caption{Local variation of the surfactant concentration $\partial \Gamma/\partial t$ (LV), surfactant convection ${\boldsymbol \nabla}_s\cdot(\Gamma {\bf v}_s)$ (C), interface expansion/compression $\Gamma {\bf n}\cdot ({\boldsymbol \nabla}_s\cdot {\bf n}){\bf v}$ (E/C), surface diffusion $\text{Pe}_s^{-1}{\boldsymbol \nabla}_s\cdot \left[{\boldsymbol \nabla}_s\Gamma/(1-\Gamma)\right]$ (SD), and adsorption/desorption flux ${\cal J}$ (A/D). The results correspond to the instant for which $F_{\textin{min}}=0.02$ ($z_{\textin{min}}$ is the vertical coordinate of the interface neck)}
\label{val2}
\end{figure}

% Surfactant conservation
Figure \ref{val2} shows the balance of the terms in the surfactant conservation equation (\ref{ad}) in the insoluble and diffusion-limited cases at the instant for which $F_{\textin{min}}=0.02$. In the insoluble case, the effect of surfactant convection (C) opposes that of interface expansion/compression (E/C). Those effects are not perfectly balanced, leading to a decrease in the surfactant concentration (LV$<0$). In the diffusion-limited case, surfactant adsorption/desorption (A/D) essentially compensates for the imbalance between surfactant convection (C) and interface expansion/compression (E/C), keeping the surfactant concentration practically constant across the interface (Fig.\ \ref{st7}b). Interestingly, surface diffusion (SD) produces a negligible effect even though the diffusion coefficient considerably increases due to the monolayer quasi-saturation ($\Gamma\simeq 1$).

\subsection{Thinning of the interface neck}
\label{bre}

% Expectation
Once the filament connecting the upper and lower liquid volumes has formed, its neck rapidly thins while the rest of the filament's shape remains practically unaltered. This disparity in time scales suggests using the time to the pinching $\tau=t_b-t$ ($t_b$ is the interface breakup time) to describe the neck filament's thinning. Neck thinning occurs on a decreasing timescale, which challenges surfactant diffusion's ability to maintain a constant surfactant concentration. One expects surfactant sorption to become negligible in the interface neck as the interface approaches its breakup. The following scaling analysis shows the validity of this expectation.

% Scaling analysis
% Convection
The surfactant surface convection term, ${\boldsymbol \nabla}_s\cdot(\Gamma {\bf v}_s)$, can be split into two contributions: $\Gamma {\boldsymbol \nabla}_s\cdot {\bf v}_s$ and ${\boldsymbol \nabla}_s \Gamma\cdot {\bf v}_s$. The order of magnitude of $\Gamma {\boldsymbol \nabla}_s\cdot {\bf v}_s$ can be obtained as follows:
\begin{equation}
\Gamma {\boldsymbol \nabla}_s\cdot {\bf v}_s\sim \Gamma w_z\sim \Gamma u_r\sim \Gamma F_{\textin{min}}^{-1}(-dF_{\textin{min}}/d\tau)\sim \Gamma\, \alpha\, \tau^{-1},
\end{equation}
where we have considered that $F_z=0$, $v_s=w$, and $\partial/\partial z=\partial/\partial s$ in the interface neck, $w_z=u_r$ due to mass conservation, $u=-dF_{\textin{min}}/d\tau$ due to the kinematic compatibility condition (\ref{int1}), and $F_{\textin{min}}\sim \tau^{\alpha}$ ($\alpha=2/3$ in the inertio-capillary asymptotic regime). 

For the inertio-capillary asymptotic regime, ${\boldsymbol \nabla}_s \Gamma\cdot {\bf v}_s$ scales as: 
\begin{equation}
{\boldsymbol \nabla}_s \Gamma\cdot {\bf v}_s\sim \Gamma_z w\sim \Delta\Gamma/\Delta z\, w\sim \Delta\Gamma\, \tau^{-2/3}\, w\sim  \Delta\Gamma\,  \tau^{-1},
\end{equation}
where $\Delta\Gamma$ is the scale of the surfactant coverage variation in the neck, and $\Delta z\sim \tau^{2/3}$ and $w\sim  \tau^{-1/3}$ are the neck characteristic axial length and velocity, respectively \citep{CCTSHHLB15}. %This term is much smaller than $\Gamma {\boldsymbol \nabla}_s\cdot {\bf v}_s$ because $\Delta\Gamma\ll\Gamma$ (Fig.\ \ref{st7}b).

% Expansion/compression
The order of magnitude of the expansion/compression term in Eq. (\ref{ad}) is
\begin{equation}
\label{mio3}
\Gamma {\bf n}\cdot ({\boldsymbol \nabla}_s\cdot {\bf n}){\bf v}\sim \Gamma F_{\textin{min}}^{-1}  (-dF_{\textin{min}}/d\tau)\sim \Gamma\, \alpha\, \tau^{-1},
\end{equation}
where we have taken into account that the curvature ${\boldsymbol \nabla}_s\cdot {\bf n}\simeq F_{\textin{min}}^{-1}$ and $v_n=u=-dF_{\textin{min}}/d\tau$ at the interface neck.

% Diffusion. 
To analyze the diffusion term, we use a frame of reference solidly moving with the interface (the surfactant boundary layer) so that the term $\partial c/\partial t$ in Eq.\ (\ref{c11}) does not need to be considered. In the interface neck, $c_z\ll c_r$, $c_{zz}\ll c_{rr}$, and $c_r/r\ll c_{rr}$ because $\delta\ll F$ (Fig.\ \ref{boun}). The balance between the dominant contributions to the convection and diffusion terms in Eq.\ (\ref{c11}) yields
\begin{equation}
\label{s0}
u'\, c_r\sim\text{Pe}^{-1}c_{rr},    
\end{equation}
where $u'\sim -(\delta/F_{\textin{min}})\, dF_{\textin{min}}/d\tau$ is the velocity normal to the interface in the interface frame of reference. These terms are of the order $u' \Delta c/\delta$ and $\text{Pe}^{-1} \Delta c/\delta^2$ in the surfactant boundary layer, where $\Delta c$ is the decrease in the surfactant concentration across the boundary layer. Since $F_{\textin{min}}\sim \tau^{\alpha}$, one obtains $u'\sim \delta\, \alpha\, \tau^{-1}$ and, therefore, $u' \Delta c/\delta\sim \alpha\, \tau^{-1} \Delta c$. Then, Eq.\ (\ref{s0}) yields 
\begin{equation}
\delta\sim [\tau/(\text{Pe}\, \alpha)]^{1/2}.
\end{equation}
This result shows how the boundary layer thins as the interface approaches its breakup. For the instant considered in Fig.\ \ref{boun}, $\tau\sim 10^{-2}$ and $\delta\sim 10^{-3}$, which agrees with the simulation result. The surfactant adsorption term can be estimated as
\begin{equation}
\label{mio2}
{\cal J}\sim \frac{\Delta c}{\delta \text{Pe}}\sim \Delta c\,  (\tau \text{Pe}/\alpha)^{-1/2}.
\end{equation}
%We have verified that the results of our scaling analysis are consistent with the values shown in Fig.\ (\ref{val2}) for $z=z_{\textin{min}}$. Specifically, the diffusion term is one order of magnitude less than that of the other terms.  

%where we have considered that $0.1\lesssim \Delta c\lesssim c_0$ (Fig.\ \ref{boun}). As shown in Fig.\ \ref{st}, $c_0=\hat{c}_0/(\hat{\Gamma}_{\infty}/R)\sim 10^3$.

\begin{figure}
\begin{center}
\resizebox{0.45\linewidth}{!}{\includegraphics{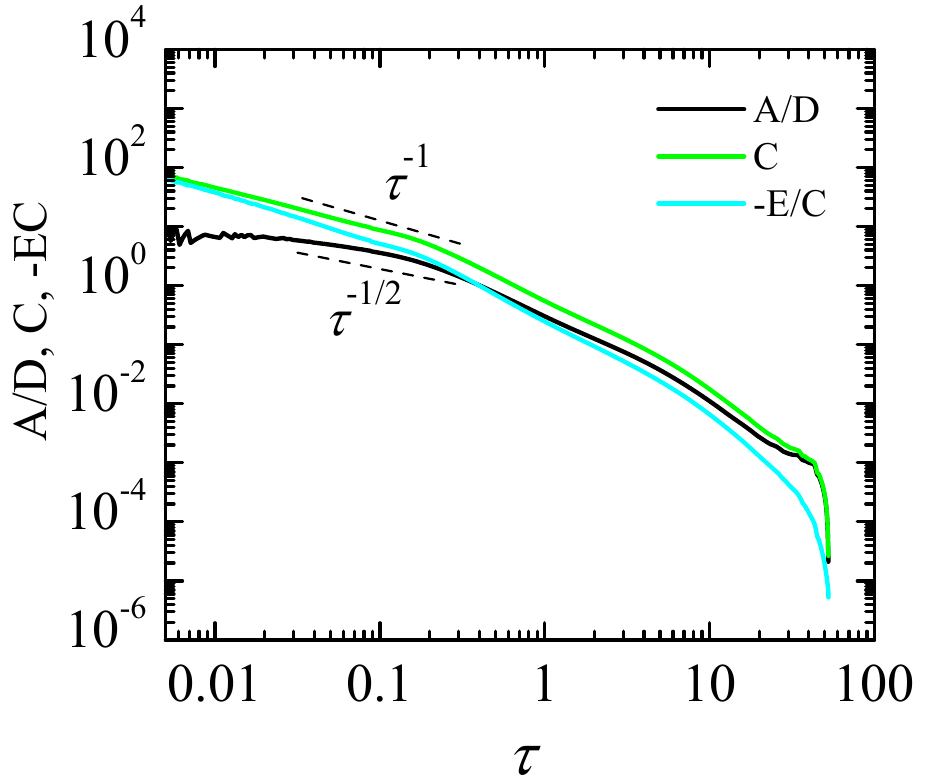}}
\end{center}
\caption{The adsoption/desorption (A/D), convection (C), and expansion/compression (E/C) terms evaluated at the interface neck as a function of the time to the pinching $\tau$ fro the diffusion-limited model.}
\label{terms}
\end{figure}

% Numerical results
The major conclusion of the above scaling analysis is that the surfactant adsorption/desorption term scales as $\tau^{-1/2}$ while the expansion/compression and convection terms scale as $\tau^{-1}$ next to the interface breakup. This implies that surfactant sorption becomes subdominant in the neck as the interface approaches its breakup. This is confirmed by the results shown in Fig.\ \ref{terms}. As predicted by the scaling analysis, the convection and expansion/contraction terms scale as $\tau$. There is an interval of time to the pinching within which the adsorption/desorption term scales as $\tau^{1/2}$. For smaller values of $\tau$, $\partial c/\partial r$ at the interface either decreases below the estimated value $\Delta c/\delta$ or $\Delta c$ decreases with $\tau$. Therefore, the term A/D deviates from the prediction of the scaling analysis. For $\tau\lesssim 0.006$ ($F_{\textin{min}}\lesssim 0.02$), small fluctuations in $\Gamma$ due to the limited spatiotemporal resolution produce large fluctuations in $c_s$ because $\Gamma\simeq 1$ (see Eq.\ (\ref{km})). This results in large fluctuations of the term A/D, which prevents us from analyzing accurately surfactant sorption in this last regime. It must be pointed out that this occurs for times to the pinching for which the sorption kinetics term becomes subdominant. Therefore, these fluctuations do not significantly affect the evolution of the rest of variables. 

\begin{figure}
\begin{center}
\resizebox{0.55\linewidth}{!}{\includegraphics{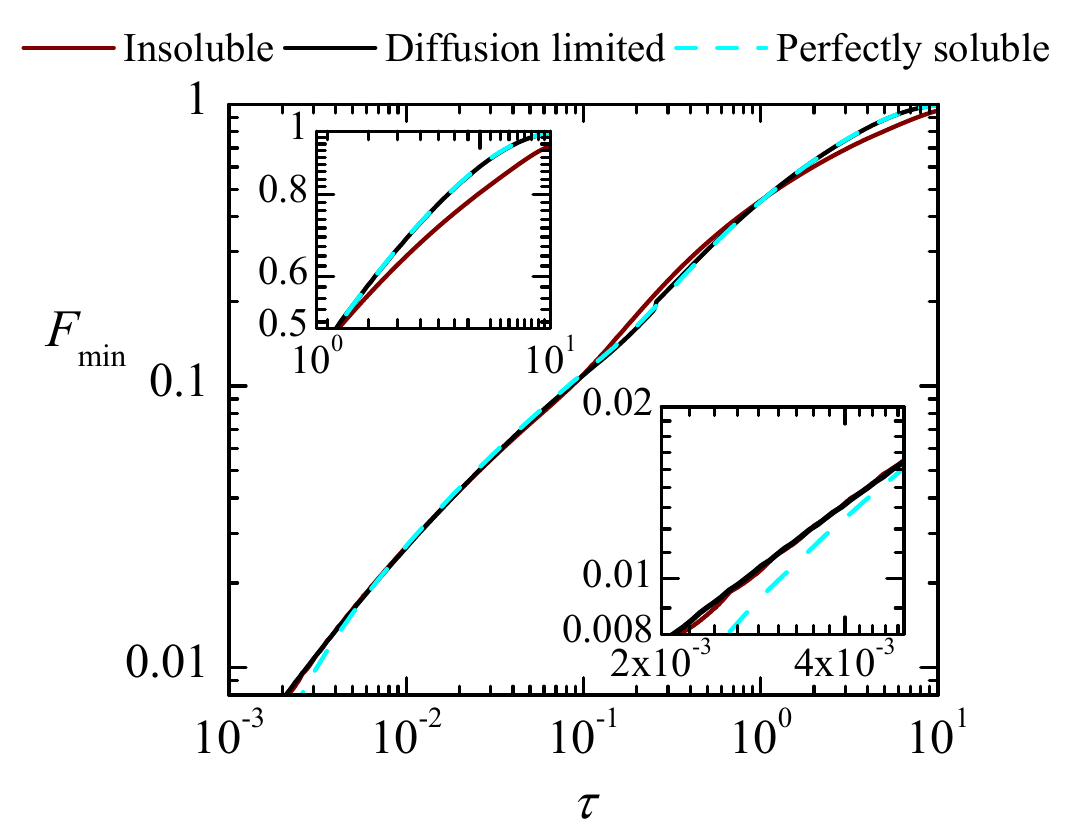}}
\end{center}
\caption{$F_{\textin{min}}$ as a function of the time to the pinching $\tau$. The insets in the upper-left and lower-right corners show $F_{\textin{min}}(\tau)$ at the beginning and at the end of the simulation, respectively.}
\label{val200}
\end{figure}

Overall, surfactant adsorption perfectly compensates for surfactant depletion due to surfactant convection at the beginning of breakup, collaborates with interface compression to balance surfactant convection at the intermediate breakup phase, and becomes subdominant for $\tau\lesssim 0.01$. Figure \ref{val200} confirms that the diffusion-limited model behaves as the perfectly soluble one for $\tau\gtrsim 0.015$ and approaches the insoluble one as $\tau\to 0$. In this limit, pinch-off occurs faster both in the diffusion-limited and insoluble case ($\tau$ is smaller for the same neck radius $F_{\textin{min}}$) because surfactant depletion near the breakup point decreases the driving capillary pressure. 

The results obtained from the three models practically coincide in the interval $10^{-2}\lesssim \tau\lesssim 10^{-1}$, indicating that $F_{\textin{min}}(\tau)$ is almost insensitive to sorption kinetics within that intermediate phase despite the significant differences between the droplet shapes (Fig.\ \ref{st6}). In other words, $F_{\textin{min}}(\tau)$ alone is not a good indicator of the surfactant monolayer dynamics.

\section{Experimental results}
\label{exp2}

% The criterion for the comparison
This section presents the experimental results. First, we compare the diffusion-limited model predictions with experimental data for Surfynol 465. In the experiment, droplet breakup begins as the unstable eigenmode spontaneously grows, making it impossible to determine the initial instant of the process from the experimental images. For this reason, we considered the simulation contour at a given instant and assigned that time to the experimental image with the same $F_{\textin{min}}$ value. The values of $t$ were assigned to the other images accordingly.

\begin{figure*}
\begin{center}
\resizebox{0.85\textwidth}{!}{\includegraphics{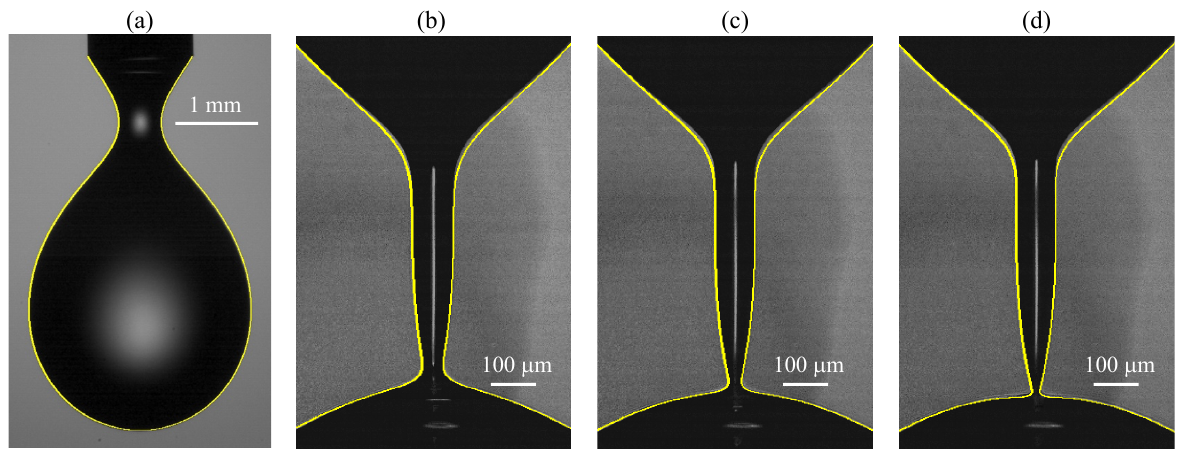}}
\end{center}
\caption{Experimental images in the presence of Surfynol 465 and numerical interface contour (yellow lines) calculated with the diffusion-limited model for $\beta=0.005$, $B=0.0954$, $\text{Oh}=6.18\times 10^{-3}$, $\Gamma_0=0.997$, $\Lambda_d=0.599$, $\text{Ma}=0.125$, $\text{Pe}_s=1.62\times 10^{4}$, and $\text{Pe}=1.62\times 10^{4}$. The instants correspond to $F_{\textin{min}}=0.4$ (a), 0.04 (b), 0.025 (c), and 0.015 (d).}
\label{val1}
\end{figure*}

\begin{figure}
\begin{center}
\resizebox{0.5\linewidth}{!}{\includegraphics{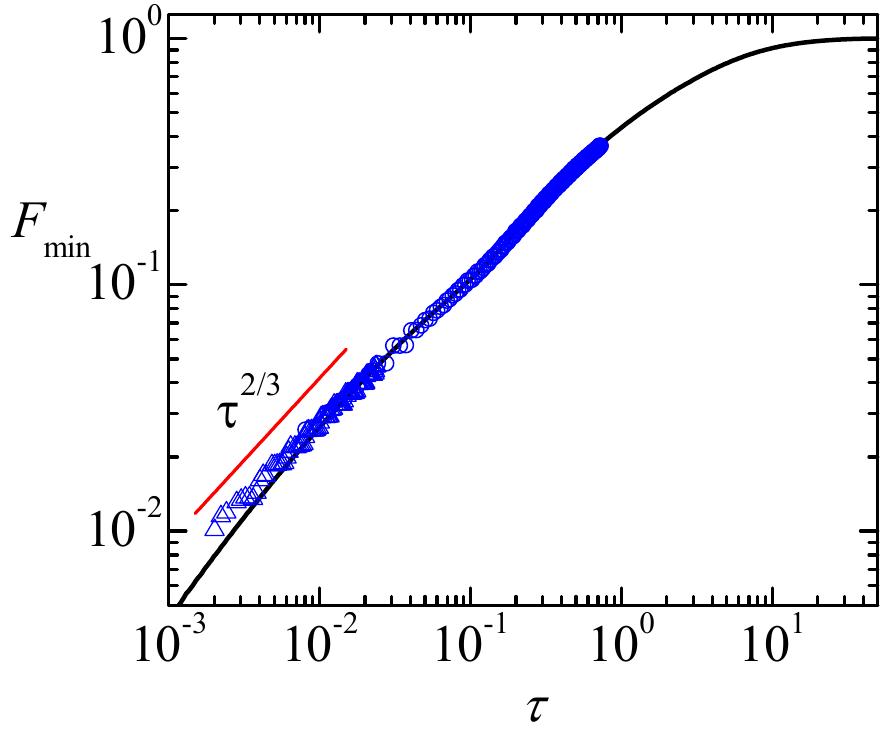}}
\end{center}
\caption{$F_{\textin{min}}$ as a function of the time to the pinching $\tau$. The symbols are the experimental results in the presence of Surfynol 465, while the line is the numerical prediction calculated with the diffusion-limited model for $\beta=0.005$, $B=0.0954$, $\text{Oh}=6.18\times 10^{-3}$, $\Gamma_0=0.997$, $\Lambda_d=1.67$, $\text{Ma}=0.125$, $\text{Pe}_s=1.62\times 10^{4}$, and $\text{Pe}=1.62\times 10^{4}$. The circles and triangles correspond to two experimental realizations with different optical magnifications. The experimental uncertainty is of the order of the symbol size.}
\label{val20}
\end{figure}

% Surfynol 465
Figures \ref{val1} and \ref{val20} show remarkable agreement between the numerical and experimental droplet shapes down to $F_{\textin{min}}\simeq 0.015$ ($\hat{F}_{\textin{min}}\simeq 9$ $\mu$m) and $\tau\simeq 0.004$ ($\tau t_{c}\simeq 10$ $\mu$s). This suggests that Surfynol 465 sorption kinetics does not constitute a significant energy barrier on that timescale. As the interface approaches breakup, this energy barrier hinders the restoration of surfactant near the pinch-off point, the driving capillary pressure increases, and neck thinning accelerates (the experimental value of $\tau$ is smaller than the numerical one for the same neck radius $F_{\textin{min}}$) (Fig.\ \ref{val20}). 

Surfynol 465 behaves as a diffusion-limited surfactant for $F_{\textin{min}}\geq 0.015$. In addition, the diffusion-limited and perfectly soluble models predict a practically identical evolution of filament shape (Figs.\ \ref{st6} and \ref{val200}). These two results, combined, allow us to conclude that Surfynol 465 practically maintains the equilibrium surface tension across the entire interface for $F_{\textin{min}}\geq 0.015$.

% The criterion for the comparison
We finish our study by comparing the droplet evolution in the presence of Surfynol 465, SDS, and Triton X-100. We chose the concentrations of the three surfactants so that they approximately correspond to the same equilibrium surface tension. Therefore, the experiments share common values of $B$, $\text{Oh}$, and $\Gamma_0$. The isotherms $\hat{\sigma}(\hat{\Gamma}/\hat{\Gamma}_{\infty})$ for the three surfactants are very similar (Fig.\ \ref{st}). This implies that the difference among their effects must be attributed essentially to the solubility parameters (sorption kinetics) $\Lambda_d$ and Bi. As explained above, we assigned the same time to the images for which $F_{\textin{min}}$ was the same. The values of $t$ were assigned to the other images accordingly. 

\begin{figure}
\begin{center}
\resizebox{0.45\textwidth}{!}{\includegraphics{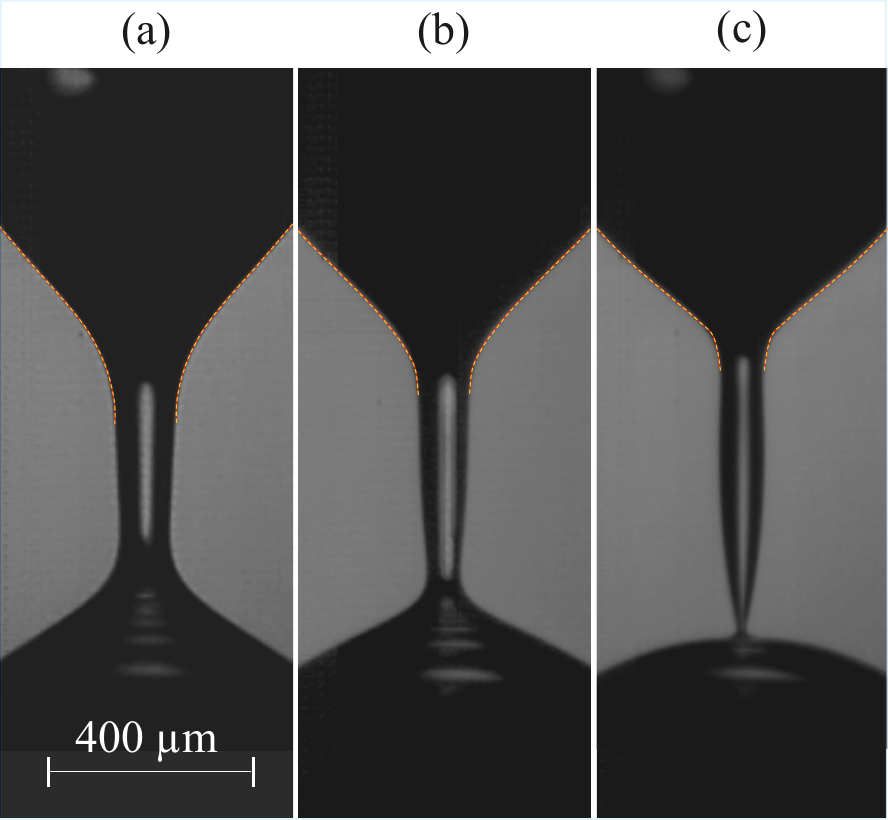}}
\end{center}
\caption{Superposition of the experimental images in the presence of Surfynol 465 and SDS for $B=0.0954$, $\text{Oh}=6.18\times 10^{-3}$, and $\Gamma_0\simeq 1$. The interface contours practically overlap. The red dashed line corresponds to Surfynol 465, while the yellow line corresponds to SDS. The instants correspond to $F_{\textin{min}}=0.103$ (a), 0.071 (b), and 0.016 (c).}
\label{val10}
\end{figure}

% Surfynol 465 and SDS
Figure \ref{val10} shows that the filament shape was practically the same for Surfynol 465 and SDS across the neck thinning. This suggests that (i) SDS can also be considered as a fast-kinetics surfactant during the breakup of a millimeter drop, as suggested by the results of the maximum bubble-pressure tensiometer (Fig.\ \ref{bmp}), and (ii) the difference between the values of $\Lambda_d$ has a negligible effect. The droplet breakup occurs on a time scale of the order of 10 ms. The results in Fig.\ \ref{bmp} suggest that the interface can be replenished with surfactant over most part of the droplet breakup, consistently with our findings. 

\begin{figure}
\begin{center}
\resizebox{0.45\textwidth}{!}{\includegraphics{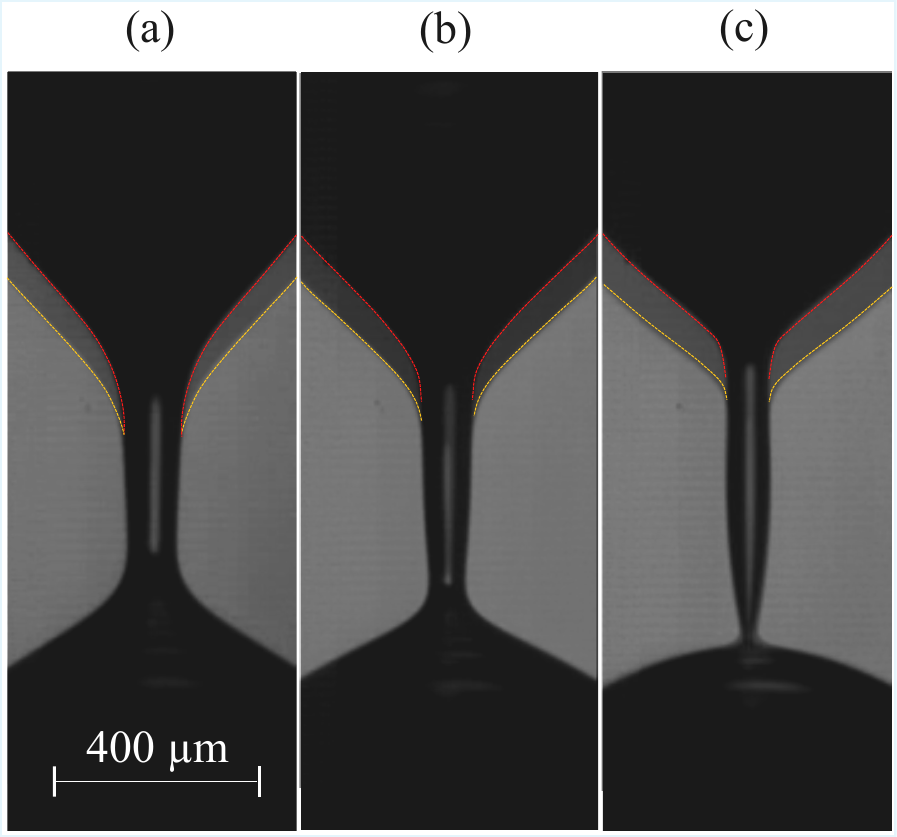}}
\end{center}
\caption{Superposition of the experimental images in the presence of Surfynol 465 and Triton X-100 for $B=0.0954$, $\text{Oh}=6.18\times 10^{-3}$, and $\Gamma_0\simeq 1$. The red dashed line corresponds to Surfynol 465, while the yellow line corresponds to Triton X-100. The instants correspond to $F_{\textin{min}}=0.103$ (a), 0.071 (b), and 0.016 (c) in the experiment with Surfynol 465.}
\label{val11}
\end{figure}

% Surfynol and Triton
Figure \ref{val11} compares the filament shape in the presence of Surfynol 465 and Triton X-100. The interface radius in the upper part of the filament was significantly larger in the presence of Triton X-100, and the length of the filament was considerably shorter. This can be attributed to an increase in surface tension due to depletion of Triton X-100, as predicted by the insoluble model (Fig.\ \ref{st7}). This depletion is caused by the Triton X-100 adsorption energy barrier, which becomes relevant in this phase of breakup. 

\begin{figure}
\begin{center}
\resizebox{0.5\linewidth}{!}{\includegraphics{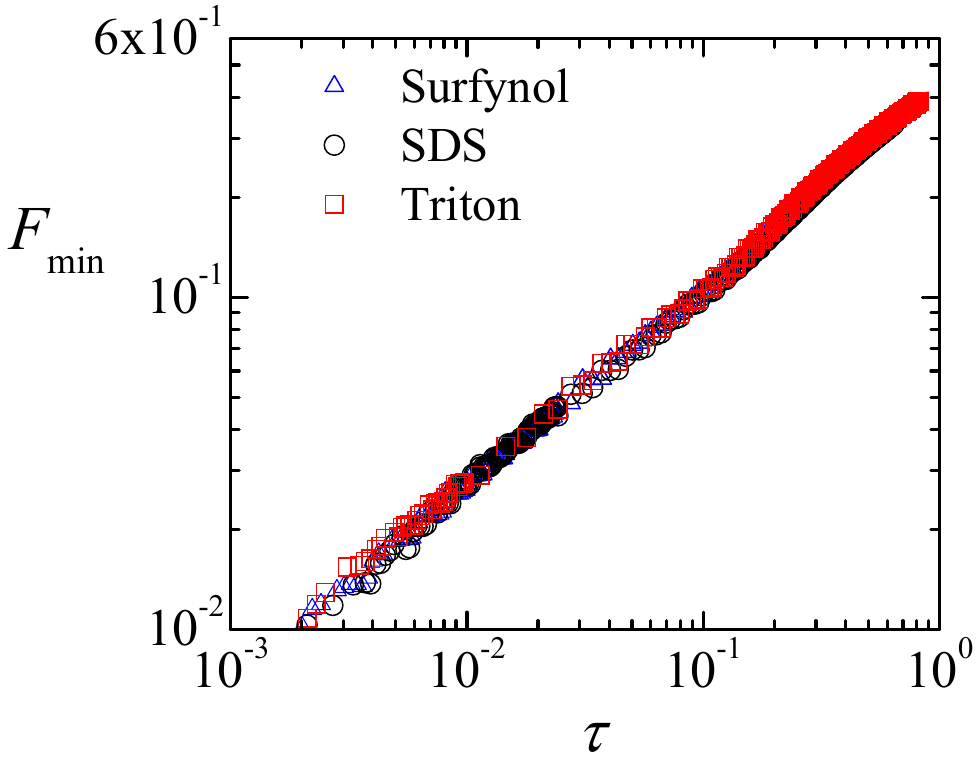}}
\end{center}
\caption{$F_{\textin{min}}$ as a function of the time to the pinching $\tau$ in the experiments with Surfynol 465, SDS, and Triton X-100 for $B=0.0954$, $\text{Oh}=6.18\times 10^{-3}$, and $\Gamma_0\simeq 1$. The experimental uncertainty is of the order of the symbol size.}
\label{val30}
\end{figure}

% Fmin
Figure \ref{val30} compares the interface neck radius versus the time to the pinching $\tau$ in the experiments with Surfynol 465, SDS, and Triton X-100. As anticipated in Sec.\ \ref{bre}, $F_{\textin{min}}(\tau)$ is not a good indicator of the surfactant monolayer dynamics because the results are practically insensitive to the surfactant sorption kinetics within the interval of $\tau$ explored in the experiments. The increase in surface tension due to surfactant depletion in the neck produces two opposing effects: an increase in the driving capillary pressure, which accelerates pinching, and a Marangoni stress that slows the interface breakup. This may explain why $F_{\textin{min}}(\tau)$ is not sensitive to the surfactant monolayer evolution,

We conclude that the most evident effect of the surfactant adsorption energy barrier is the swelling of the upper section of the liquid filament connecting the two liquid volumes prior to interface pinch-off, or, equivalently, the shortening of that filament. In other words, comparing the filament length (or, equivalently, $z_{\textin{min}}(\tau)$) with that of a clean interface for the same equilibrium surface tension provides a measure of the surfactant adsorption rate.

\section{Concluding remarks}
\label{conclusions}

% Results
We have investigated numerically and experimentally the breakup of a pendant droplet containing a soluble surfactant in the diffusion-limited (fast-kinetics) regime. A thin liquid filament connecting the upper and lower liquid volumes forms on a timescale of order of the inertio-capillary time. Convective surfactant transport associated with filament thinning enhances surfactant transfer toward the interface, such that diffusion does not significantly hinder adsorption over most of the breakup process. As a result, the surface tension remains nearly uniform, and the dynamics closely follow those of a clean droplet with the same equilibrium surface tension. Diffusion becomes a limiting factor only in the immediate vicinity of pinch-off, where the neck radius evolution $F_{\textin{min}}(\tau)$ approaches that of the insoluble limit. Significant differences between the diffusion-limited and insoluble models are observed during filament formation. In the insoluble case, surfactant depletion increases surface tension and induces Marangoni stress, altering the filament shape. 

Experiments with millimeter-sized droplets containing Surfynol 465 are in excellent agreement with the diffusion-limited model, with no adjustable parameters. SDS exhibits similar behavior, indicating that both surfactants effectively maintain a constant surface tension during most of the breakup. In contrast, a slow-kinetics surfactant such as Triton X-100 leads to markedly different dynamics. The more noticeable footprint of the surfactant adsorption energy barrier is the bulging of the upper part of the liquid filament connecting the upper and lower liquid volumes before the interface pinching. Comparing the filament length (or, equivalently, $z_{\textin{min}}(\tau)$) to that of a clean interface with the same surface tension allows one to assess the rate of surfactant adsorption.

% Droplet size
It must be emphasized that our analysis is restricted to the common case of a millimeter-sized droplet, whose breakup occurs on a timescale of approximately 10 ms. SDS may essentially behave as an insoluble surfactant for submillimeter droplets with breakup times lower than 1 ms \citep{PRHEM20}. Recent experiments have shown that Surfynol 465 critically influences cavity dynamics following microjet impact on a liquid pool on sub-millisecond timescales \citep{FVMGF26}. Conversely, the cavity collapse pathway in the presence of SDS is nearly indistinguishable from that in pure water, suggesting a negligible surfactant adsorption on sub-millisecond timescales. Based on these results, one expects to find significant differences in the effects of Surfynol 465 and SDS on droplets with radii of order 0.1 mm or less.

%\section*{Acknowledgement}
\begin{bmhead}[Acknowledgement]
We gratefully acknowledge support from the Spanish Ministry of Science and Innovation (MCIN) (grant no. PID2022-140951OB-C21 and PID2022-140951OB-C22/ AEI/10.13039/501100011033/ FEDER, UE), Junta de Extremadura (grant nos. GR21091 and IB24089). DF-M acknowledges grant PREP2022-000205 funded by MICIU/AEI /10.13039/501100011033 and ESF+. Generative artificial intelligence tools have been used for revising the text.
\end{bmhead}

The authors report no conflict of interest.

\end{document}